\documentstyle{article}

         \def\ba{\begin{array}}
         \def\ea{\end{array}}
         \def\be{\begin{equation}}
         \def\bea{\begin{eqnarray}}
         \def\eea{\end{eqnarray}}
         \def\ee{\end{equation}}
         
         \def\l{{\lambda}}

         \def\DD{\Delta}
         \def\D{\delta}

         \def\bq{\begin{eqnarray}}
         \def\eq{\end{eqnarray}}
         
\begin{document}
\begin{titlepage}
\hfill
\vbox{
    \halign{#\hfil         \cr
           hep-th/9807034\cr
           IPM/P-98/16\cr
           } 
      }  
\vspace*{3mm}
\begin{center}
\hfill
\vskip 5 mm
\leftline{ \Large \bf
           Logarithmic conformal field theories and AdS correspondence}
\vskip 1cm
\leftline{\bf A. M. Ghezelbash$^{1,3,*}$, M. Khorrami$^{2,3}$,
              A. Aghamohammadi$^{1,3}$
              }
\vskip .5cm
{\it
  \leftline{ $^1$ Department of Physics, Alzahra University,
             Tehran 19834, Iran. }
  \leftline{ $^2$ Department of Physics, Sharif University of Technology
             P.O. Box 11365-9161, Tehran, Iran. }
  \leftline{ $^3$ Institute for Studies in Theoretical Physics and
            Mathematics, P.O.Box  5531, Tehran 19395, Iran. }
  \leftline{ $^*$ amasoud@netware2.ipm.ac.ir}
  }

\vskip 10 mm

\end{center}
\begin{abstract}
\vskip 5 mm

We generalize the Maldacena correspondence to the logarithmic conformal field
theories. We study the correspondence between field theories in
(d+1)-dimensional AdS space and the d-dimensional logarithmic conformal field
theories in the boundary of $AdS_{d+1}$. Using this correspondence, we get the
n-point functions of the corresponding logarithmic conformal field theory in
d-dimensions.
\end{abstract}
\end{titlepage}
\newpage

\newpage
\section{\bf Introduction}
Various aspects of the correspondence between field theories in
$(d+1)$--dimensional Anti de  Sitter space (AdS) and $d$--dimensional
conformal field theories (CFT's) has been studied in the last few months. An
important example is the conjectured correspondence between the large $N$
limits of certain conformal field theories in $d$--dimensions and
supergravity on the product of a $(d+1)$--dimensional AdS space with a compact
manifold \cite{Ma}. This suggested correspondence was made more precise in
\cite{GKP,FF,W}.

The general correspondence between a theory on an AdS and a conformal theory
on the boundary of AdS is the following. Consider the partition function of
a field theory on AdS, subjected to the constraint
\be\label{1}
\phi_{\vert_{\partial AdS}}=\phi_0,
\ee
that is
\be
Z_{\rm AdS}(\phi_0)=\int_{\phi_0}{\rm D}\phi\exp [iS(\phi )],
\ee
where the functional integration is over configurations satisfying
(\ref{1}). It is well known that the symmetry algebra of a $(d+1)$--dimensional
AdS is O($d$,2), which is the same as the conformal algebra on a $d$--
dimensional Minkowski space. From this, it is seen that
$Z_{\rm AdS}(\phi_0)$ is invariant under conformal transformations, and this
is the root of the analogy between theories on AdS and conformal theories on
$\partial{\rm AdS}$. In fact, if
\be
\phi\to O\phi,
\ee
is a space--time symmetry of the theory on AdS, it is seen that
\bea
Z_{\rm AdS}(O^{-1}\phi_0)&=&\int_{\phi_0}{\rm D}(O\phi )\exp [iS(O\phi )],
\cr &=&\int_{\phi_0}{\rm D}\phi\exp [iS(\phi )],
\eea
which means that
\be
Z_{\rm AdS}(O^{-1}\phi_0)=Z_{\rm AdS}(\phi_0).
\ee
So, one can use $Z_{\rm AdS}(\phi_0)$ as the generating function of a
conformally invariant theory on the boundary of AdS, with $\phi_0$ as the
current.

The notion of the boundary of an AdS needs, however, some care. An
AdS$_{d+1}$ is a $(d+1)$--dimensional hypersurface in $R^{d+2}$ equiped with
the metric
\be
\eta ={\rm diag}(-1, -1, 1, \cdots ,1).
\ee
The equation of this hypersurface is
\be
(X^{-1})^2+(X^0)^2-\sum_{i=1}^d (X^i)^2=1.
\ee
This is a connected hyperboloid in $R^{d+2}$. By boundary, it is meant the
points satisfying
\be\label{8}
\sum_{i=1}^d (X^i)^2=R^2\to\infty .
\ee
The above set (for fixed $R$) is the $d$--dimensional sapce
$S\times S_{d-1}$. As $R\to\infty$, this space tends to a space which is
locally $R^d$, since the radii of $S$ and $S_{d-1}$ are $\sqrt{R^2+1}$ and
$R$, respectively.

Using the coordinate transformation
\bea
X^i&=&{x^i\over{x^d}},\qquad i\ne d,-1\cr x^d&=&{1\over{X^{-1}+X^d}},
\eea
the length element is seen to take the form
\be\label{10}
{\rm d}s^2={{-({\rm d}x^0)^2+\sum_{i=1}^d({\rm d}x^i)^2}\over{(x^d)^2}}.
\ee
Almost all of the boundary is now contained in $x^d=0$. There are, however,
points satisfying (\ref{8}), but not $X^{-1}+X^d\to\infty$.These points
serve as compactifiers of $R^d$ to $S\times S_{d-1}$. We use the length
element (\ref{10}), and use $x^d=0$ as the boundary of AdS. In this case,
the boundary will be a $d$--dimensional Minkowski space.

The above--mentioned correspondence have been studied for various cases,
e. g. a free massive scalar field and a free U(1) gauge theory \cite{W}, an
interacting massive scalar field theory \cite{MV}, free massive spinor
field theory \cite{HS}, and interacting massive spinor-scalar field theory
\cite{GKPF}. Our aim in this article is to study the the
correspondence between theories on AdS spaces and logarithmic conformal
field theories (LCFT's). It has been shown by Gurarie \cite{Gu}, that conformal
field
theories whose correlation functions exhibit logarithmic behaviour,
can be consistently defined. It is shown that if in the OPE of two local
fields, there exist at least two fields with the same conformal dimension,
one may find some special operators, known as logarithmic
operators. As discussed in \cite{Gu}, these operators with the ordinary
operators form the basis of a Jordan cell for the operators $L_i$ (the
generators of the conformal algebra).
In some interesting physical theories, for example dynamics
of polymers \cite{Sa}, the WZNW model on the $GL(1,1)$ super-group \cite{RS},
WZNW models at level 0 \cite{KM,CKLT,KL}, percolation \cite{Ca},
the Haldane-Rezayi quantum Hall state\cite{GFN}, and edge excitation in
fractional quantum Hall effect \cite{WWH}, one can naturally find
logarithmic terms in correlators. Recently the role of logarithmic operators
have been considered in study of some physical problems such as
2D--magnetohydrodynamic turbulence \cite{rr1,rr2,rr3}, 2D--turbulence
\cite{flo,rahim1}, $c_{p,1}$ models \cite{GK,fl},
gravitationally dressed CFT's \cite{BK}, and some critical disordered
models \cite{CKT,MS}.
Logarithmic conformal field theories for the supersymmetric case \cite{CKLT,MAG}
and for
$d$--dimensional case ($d>2$)
have also been studied \cite{GhK}.
The basic properties of logarithmic operators are that
they form a part of the basis of the Jordan cell
for $L_i$'s, generators of the Virasoro algebra, and in the correlator of
such fields there  is a logarithmic singularity \cite{Gu,CKT}.
In \cite{RAK} assuming conformal invariance  two-- and
three--point functions for the case of one or more logarithmic fields
in a block, and one or more  sets of logarithmic fields
have been explicitly calculated. Regarding logarithmic fields
{\it formally} as derivatives of ordinary fields with respect to their
conformal dimension,  $n$--point functions containing logarithmic fields
have been calculated in terms of those of ordinary fields.
These have been done when conformal weights belong to a discrete set.
In \cite{KAR}, logarithmic conformal field theories with continuous weights
have been considered. It is shown in \cite{RAK} that if the set of weights
is discrete, when the Jordan cell for $L_i$ is two dimensional,
there are two fields ${\cal O}$ and ${\cal O}'$, with the following
two-point functions
\be
   \langle{\cal O}({\bf x}){\cal O}({\bf y}) \rangle =0,
\ee
\be
   \langle{\cal O}({\bf x}){\cal O}'({\bf y}) \rangle =
   \frac{c}{|{\bf x}-{\bf y}|^{2\Delta}},
\ee
\be
   \langle{\cal O}'({\bf x}){\cal O}'({\bf y}) \rangle =
   \frac{1}{|{\bf x}-{\bf y}|^{2\Delta}}
   \left( c'-2c\ln |{\bf x}-{\bf y}|
   \right).
\ee

In \cite{KA} another type of derivation is also introduced. The main idea of
this construction is based on the {\it formal} derivation of the entities of
the original system with respect to a parameter, which may or may not
explicitly appear in the original theory. One way of viewing this is through
the concept of contraction: consider two systems with parameters $\l$ and
$\l +\D$. These two systems are independent to each other. One can write an
action as the difference of the actions of the two system divided by $\D$ to
describe both systems. One can use one of these degrees of freedom and the
difference of them divided by $\D$ a new set of variables. This system,
however, is equivalent to two copies of the original system. But if one lets
$\D$ tend to zero, a well--defined theory of double number of variables is
obtained, which no longer can be decomposed to two independent parts. One
can, however, solve this theory in terms of the solution of the original
theory. This procedure is nothing but a contraction. It has been shown in
\cite{KAA} that any symmetry, and any constant of motion of the original
theory, results in a symmetry and a constant of motion of the derived one
and  any theory derived from an integrable theory is integrable. At last, it
has also been shown that this technique is applicable to classical field
theories as well. This technique is applicable to quantum systems as well.
Here, however, a novel property arises: the derived quantum theory is
{\it almost classical}; that is, in the derived theory there are only
one--loop quantum corrections to the classical action \cite{KAA}. Using this
property, one can calculate all of the Green functions of the derived theory
exactly, even though this may be not the case for the original theory.

There may arise some interesting questions, e.g. \newline
{\it is there any
correspondence between  field theories in $(d+1)$--dimensional AdS space and
$d$--dimensional logarithmic conformal field theories}? \newline
This may shed light on
logarithmic conformal field theories and the AdS/LCFT correspondence. We
show that by a suitable choice of action in AdS$_{d+1}$, which in the
simplest case depends on the two massive scalar fields $\phi$ and $\psi$,
one gets a LCFT on the $\partial$AdS. In general, using any field theory on
AdS, which corresponds to a CFT on $\partial$AdS, one can systematically
construct other theories on AdS corresponding to LCFT's on $\partial$AdS.

\section{\bf Free theories and two--point functions}

Consider the following action,
\be
\label{ac}
 S(\phi ,\psi )={1\over 2}\int {\rm d}^{d+1}x\; \sqrt{|g|} (-2\nabla\phi
 \cdot\nabla\psi -2m^2\phi\psi -\mu^2\phi^2),
\ee
where
\be
m^2:=\DD (\DD -d),\qquad \mu^2:=2\DD -d
\ee
The equation of motion for the fields $\phi$ and $\psi$ are
\be
\label{phi}
  (\nabla^2 - m^2) \phi =0,
\ee
\be
\label{psi}
  (\nabla^2 - m^2) \psi -\mu^2\phi =0.
\ee
It is easily seen that if $\phi$ is a solution of (\ref{phi}),
differentiating it with respect to $\DD$ yields a solution of (\ref{psi})
for $\psi$. The Dirichlet Green function for this system satisfies
\be\label{21}
\pmatrix{\nabla^2-m^2&0\cr -\mu^2&\nabla^2-m^2\cr}{\cal G}(x, y)=
\delta (x, y)\pmatrix{1&0\cr 0&1\cr},
\ee
together with the boundary condition
\be\label{22}
{\cal G}(x, y)_{|_{x\in\partial{\rm AdS}}}=0.
\ee
It is easy to see that
\be
{\cal G}=\pmatrix{G&0\cr G'&G\cr}
\ee
satisfies (\ref{21}) and ({\ref{22}), provided
\be
(\nabla^2-m^2)G(x,y)=\delta (x,y),
\ee
\be
G(x, y)_{|_{x\in\partial{\rm AdS}}}=0,
\ee
and
\be
G'={{\partial G}\over{\partial\DD}}.
\ee

Now use this Green function to solve the Dirichlet problem for $\phi$ and
$\psi$ in AdS. As the metric diverges on the boundary, this problem needs
some care. One should first solve the Dirichlet problem for the boundary at
$x^d=\epsilon$, and then let $\epsilon\to 0$, defining suitably scaled
fields in that limits. This has been done in \cite{MV}, with the result
\be
G(x,y)_{|_{y^d=\epsilon\to 0}}=-c\epsilon^\Delta
\left({{x^d}\over{(x^d)^2+|{\bf x}-{\bf y}|^2}}\right)^{\DD},
\ee
and
\be
\phi (x)={2c\DD}\epsilon^{\DD -d}\int_{y^d=\epsilon}{\rm d}^dy\;
\phi({\bf y},\epsilon )
\left({{x^d}\over{(x^d)^2+|{\bf x}-{\bf y}|^2}}\right)^{\DD},
\ee
where
\be
\alpha :=\DD-{d\over 2},
\ee
and
\be
c:={{\Gamma (\DD)}\over{2\pi^{d/2}\Gamma (\alpha +1)}}.
\ee
Boldface letters denote coordinates of the boundary ($x^0$ to $x^{d-1}$),
and
\be
|{\bf x}|^2:=-(x^0)^2+\sum_{i=1}^{d-1}(x^i)^2.
\ee
Note that the solutions for $G$ and $\phi$ are the same as those found in
\cite{MV}, since the equation of motion for $\phi$ is the same as that found
in \cite{MV}. Similarly,
\bea
\psi (x)={2c\DD}\epsilon^{\DD -d}\int_{y^d=\epsilon}{\rm d}^dy\;
&\Big[&\psi({\bf y},\epsilon)+\phi({\bf y},\epsilon )\ln\epsilon +
\phi({\bf y},\epsilon ){{\partial\ln (2c\DD)}\over{\partial\DD}}\cr
&&+\phi({\bf y},\epsilon )\ln
\left({{x^d}\over{(x^d)^2+|{\bf x}-{\bf y}|^2}}\right)\Big]
\left({{x^d}\over{(x^d)^2+|{\bf x}-{\bf y}|^2}}\right)^{\DD}.
\eea
Now, defining
\be
\phi_0({\bf x}):=\lim_{\epsilon\to 0}\DD\epsilon^{\DD -d}
\phi ({\bf x},\epsilon ),
\ee
and
\be
\psi_0({\bf x}):=\lim_{\epsilon\to 0}\DD\epsilon^{\DD -d}[
\psi ({\bf x},\epsilon )+\phi ({\bf x},\epsilon )(\ln\epsilon +1/\DD )],
\ee
we have
\be
\pmatrix{\phi(x)\cr \psi(x)\cr}=\int {\rm d}^dy
\pmatrix{1&0\cr \ln\left({{x^d}\over{(x^d)^2+|{\bf x}-{\bf y}|^2}}\right)&1}
\pmatrix{\phi_0({\bf y})\cr \psi_0({\bf y})\cr}
\left({{x^d}\over{(x^d)^2+|{\bf x}-{\bf y}|^2}}\right)^{\DD}.
\ee
The action for this classical configuration is (using an integration by
parts)
\be
S_{\rm cl.}(\phi_0,\psi_0)={1\over 2}\lim_{\epsilon\to 0}\epsilon^{1-d}
\int {\rm d}^dy\left[\phi ({\bf y},\epsilon){{\partial\psi ({\bf y},
\epsilon)}\over{\partial x^d}}+\psi ({\bf y},\epsilon){{\partial
\phi ({\bf y},\epsilon)}\over{\partial x^d}}\right] .
\ee
Using the definitions of $\phi_0$ and $\psi_0$, the classical action becomes
\bea\label{cac}
S_{\rm cl.}(\phi_0,\psi_0)&=&{1\over 2}\int{{{\rm d}^dx\;{\rm d}^dy}
\over{|{\bf x}-{\bf y}|^{2\DD}}}\bigg[ 2c\phi_0({\bf x})\psi_0({\bf y})\cr
&&+\left({{\partial c}\over{\partial\DD}}-2c\ln |{\bf x}-{\bf y}|\right)
\phi_0({\bf x})\phi_0({\bf y})\bigg] .
\eea
As the theory we began with is a free theory, we have
\be
W(\phi_0,\psi_0)=S_{\rm cl.}(\phi_0,\psi_0).
\ee
where $W$ is the generating function of connected diagrams. From this, one
can calculate the Green functions of the operators ${\cal O}$ and
${\cal O}'$, their corresponding currents are $\psi_0$ and $\phi_0$,
respectively:
\be\label{41}
   \langle{\cal O}({\bf x}){\cal O}({\bf y}) \rangle =0,
\ee
\be\label{42}
   \langle{\cal O}({\bf x}){\cal O}'({\bf y}) \rangle =
   \frac{c}{|{\bf x}-{\bf y}|^{2\Delta}},
\ee
and
\be\label{43}
   \langle{\cal O}'({\bf x}){\cal O}'({\bf y}) \rangle =
   \frac{1}{|{\bf x}-{\bf y}|^{2\Delta}}
   \left( c'-2c\ln |{\bf x}-{\bf y}|
   \right).
\ee
where
\be
c':=\frac{\partial c}{\partial \Delta}.
\ee
It is interesting to compare these results with the results of a free scalar
field on AdS. It is not difficult to check that the action (\ref{ac}) and
the classical action (\ref{cac}) are formal derivatives of those
corresponding to the free scalar theory with
respect to $\DD$, in the sense of \cite{KA} and \cite{KAA}. One can also see
that the two--point functions (\ref{41}) to (\ref{43}) are those of a
two dimensional Jordanian block, where the operator ${\cal O}'$ can be
interpreted as a formal derivative of ${\cal O}$, in the sense of [27--29].
We have thus obtained a link between these two kinds of derivations: A
theory on AdS induces a conformal field theory on $\partial$AdS. The formal
derivative of the first theory induces a logarithmic conformal field theory
on $\partial$AdS, the fields of which are formal derivatives of those of
the initial conformal field theory.

In our case, we begin with the action
\be
 S_0(\phi )={1\over 2}\int {\rm d}^{d+1}x\; \sqrt{|g|} (-\nabla\phi
 \cdot\nabla\phi -m^2\phi ^2).
\ee
This theory leads to a conformal field theory on $\partial$AdS \cite{MV}.
Then, we {\it formally} differentiate this action with respect to some
parameter $\zeta$, using
\be
\frac{\partial \phi}{\partial \zeta}=:\tilde \psi,
\ee
and
\be
\frac{\partial m^2}{\partial \zeta}=:\widetilde {\mu ^2}.
\ee
Now, as
\be
\frac{\partial }{\partial \zeta}=\frac{\partial \DD }{\partial \zeta}
\frac{\partial }{\partial \DD},
\ee
one can elinimate $\frac{\partial \DD }{\partial \zeta}$ from all {\it formal}
differentiations. For example,
\be
\psi=\frac{\partial \phi}{\partial \DD},
\ee
and
\be
\mu ^2=\frac{\partial m^2}{\partial \DD}.
\ee
This shows that taking $\mu ^2$ as the derivative of $m^2$ with respect to
$\DD $ does not result in any loss of generality: taking any arbitrary
parameter $\widetilde {\mu ^2}$, one can write it as,
\be
\widetilde {\mu ^2}=\mu ^2\frac{\partial \DD}{\partial \zeta},
\ee
and absorb $\frac{\partial \DD}{\partial \zeta}$ in $\psi$ as,
\be
\tilde \psi=\psi \frac{\partial \DD}{\partial \zeta}.
\ee
One can then directly use the results of \cite{MV} and use a {\it formal}
differentiation with respect to $\DD $ to arrive at (\ref{cac}).

\section{\bf Interactions and n-point functions}
Let us consider the following interaction between the fields $\phi$ and
$\psi$,
\be \label{INT}
S_I'(\phi ,\psi )=\int{\rm d}^{d+1}x\sqrt{|g|}
\sum _{n\geq 3}\frac{\phi ^{n-1}(x)}{n!}(\lambda '_n\phi (x)+n\lambda
_n\psi(x)).
\ee
The total action $S_0+S_I'$ yields the equations of motions,
\bea \label{EQU}
  (\nabla^2 - m^2) \phi &=&\sum _{n\geq 3}\frac{\lambda _n}{(n-1)!}
\phi ^{n-1},\nonumber\\
  (\nabla^2 - m^2) \psi -\mu ^2\Phi &=&\sum _{n\geq 3}\frac{\phi ^{n-2}}
  {(n-1)!}(\lambda '_n\phi +(n-1)\lambda _n\psi),
\eea
Using the Green functions $G(x,y)$ and $G'(x,y)$ introduced before, the
classical fields $\phi $ and $\psi $ satisfying the equations (\ref{EQU})
and a Dirichlet boundary condition on $\partial$AdS, satisfy the
integral equations
\bea\label{CLAS}
\pmatrix{\phi(x)\cr \psi(x)\cr}&=&\int {\rm d}^dy
\pmatrix{1&0\cr \ln\left({{x^d}\over{(x^d)^2+|{\bf x}-{\bf y}|^2}}\right)&1}
\pmatrix{\phi_0({\bf y})\cr \psi_0({\bf y})\cr}
\left({{x^d}\over{(x^d)^2+|{\bf x}-{\bf y}|^2}}\right)^{\DD}\cr
&&\cr
&&+\int {\rm d}^{d+1}y\;\sqrt{|g|}\pmatrix{G&0\cr G'&G\cr}
\sum _{n\geq 3}\pmatrix{\frac{\lambda _n}{(n-1)!}\phi ^{n-1}\cr
\frac{\phi ^{n-2}}{(n-1)!}(\lambda '_n\phi +(n-1)\lambda _n\psi)\cr}.
\eea
Now, substituting the classical solution (\ref{CLAS})
into (\ref{INT}), one obtains, to first order in $\lambda$'s,
\bea \label{SINT1}
S_I^{(1)}(\phi _0,\psi _0)=-\sum _{n\geq 3}\frac{c^n}{n!}\int
{\rm d}^dy_1\cdots{\rm d}^dy_n&\{&\phi _0({\bf y}_1)\cdots\phi _0({\bf y}_n)
I_n({\bf y}_1,\cdots ,{\bf y}_n)\cr
&&+n \phi _0({\bf y}_1)\cdots\phi _0({\bf y}_{n-1})\psi _0({\bf y}_n)
J_n({\bf y}_1,\cdots,{\bf y}_n)\},
\eea
where
\bea\label{I}
I_n&=&\int{\rm d}^{d+1}x\;\sqrt{|g|}\frac{(x^d)^{-(d+1)+n\Delta}}{
\{[(x^d)^2+|{\bf x}-{\bf y}_1|^2]\cdots [(x^d)^2+|{\bf x}-{\bf y}_n|^2]\}
^\Delta}\cr
&&\times\{\lambda '_n+\lambda _n(n\frac{c'}{c}+\ln
\frac{(x^d)^n}{\{[(x^d)^2+|{\bf x}-{\bf y}_1|^2]\cdots
[(x^d)^2+|{\bf x}-{\bf y}_n|^2]\}})\},
\eea
and
\be \label{J}
J_n=\lambda_n\int{\rm d}^{d+1}x\;\sqrt{|g|}\frac{(x^d)^{-(d+1)+n\Delta}}{
\{[(x^d)^2+|{\bf x}-{\bf y}_1|^2]\cdots [(x^d)^2+|{\bf x}-{\bf y}_n|^2]\}
^\Delta}.
\ee
Note that there is no first order contribution from the free action. We can
now read off the connected part of the tree level $n$--point functions
($n\geq 3$) of the operators ${\cal O}$ and ${\cal O}'$. From
(\ref{SINT1}), we have
\bea \label{NPOINT}
\langle{\cal O}'({\bf x}_1)\cdots{\cal O}'({\bf x}_n)\rangle&=&
-c^nI_n({\bf x}_1,\cdots ,{\bf x}_n),\cr
\langle{\cal O}'({\bf x}_1)\cdots
{\cal O}'({\bf x}_{n-1}){\cal O}({\bf x}_n)
\rangle&=&-c^nJ_n({\bf x}_1,\cdots ,{\bf x}_n).
\eea
The other correlation functions containing more than one
${\cal O}$'s vanish. For example, in the case $n=3$ one has
\bea \label{TTHREE}
\langle{\cal O}'({\bf x}_1){\cal O}'({\bf x}_2){\cal O}'({\bf x}_3)\rangle
&=&-c^3I_3,\cr
\langle{\cal O}'({\bf x}_1){\cal O}'({\bf x}_2){\cal O}({\bf x}_3)\rangle
&=&-c^3J_3,\cr
\langle{\cal O}'({\bf x}_1){\cal O}({\bf x}_2){\cal O}({\bf x}_3)\rangle
&=& 0,\cr
\langle{\cal O}({\bf x}_1){\cal O}({\bf x}_2){\cal O}({\bf x}_3)\rangle
&=&0.
\eea
Using a Feynman parametrization in (\ref{I}) and (\ref{J}), one gets
\bea \label{THREE}
\langle{\cal O}'({\bf x}_1){\cal O}'({\bf x}_2){\cal O}({\bf x}_3)\rangle
&=&-\frac{\lambda _3\Gamma(\frac{\Delta}{2}+\alpha)}{2\pi ^d}
\left[\frac{\Gamma(\frac{\Delta}{2})}{\Gamma(\alpha)}\right]^3
\frac{1}{|{\bf x}_1-{\bf x}_2|^{\Delta}|{\bf x}_1-{\bf x}_3|^{\Delta}
|{\bf x}_2-{\bf x}_3|^{\Delta}},\cr
\langle{\cal O}'({\bf x}_1){\cal O}'({\bf x}_2){\cal O'}({\bf x}_3)\rangle
&=&{{\partial}\over{\partial\Delta}}
\langle{\cal O}'({\bf x}_1){\cal O}'({\bf x}_2){\cal O}({\bf x}_3)\rangle .
\eea
This shows the logarithmic behaviour of
$\langle{\cal O}'{\cal O}'{\cal O}'\rangle$
and the usual scaling behaviour of correlation function
$\langle{\cal O}'{\cal O}'{\cal O}\rangle$.
To get more non-vanishing correlation functions,
other forms for $S_I$ must be chosen. For example, if we take the following
form for $S_I$,
\bea \label{SINT2}
S_I''(\phi,\psi )&=&\frac{\partial}{\partial\Delta}S_I'\cr
&=&\int{\rm d}^{d+1}x\sqrt{|g|}
\sum _{n\geq 3}\frac{\phi ^{n-2}(x)}{n!}[\lambda ''_n\phi ^2(x)+2n\lambda '
_n\phi(x)\psi(x)+n(n-1)\lambda _n\psi^2(x)],
\eea
then the structure of correlation functions is as follows.
$\langle{\cal O}'({\bf x}_1)\cdots{\cal O}'({\bf x}_n)\rangle$ and
$\langle{\cal O}'({\bf x}_1)\cdots{\cal O}'({\bf x}_{n-1})
{\cal O}({\bf x}_n)\rangle$
have the logarithmic structure, and
$\langle{\cal O}'({\bf x}_1)\cdots{\cal O}'({\bf x}_{n-2})
{\cal O}({\bf x}_{n-1}){\cal O}({\bf x}_n)\rangle$ has the scaling
structure. Moreover, correlation functions containing more than two
${\cal O}$'s vanish.
In general, taking the following action of interacting fields $\phi$ and
$\psi$,
\be \label{GENERAL}
S_I(\phi ,\psi )=\int{\rm d}^{d+1}x\sqrt{|g|}
\sum _{n\geq 3}\sum _{i,j\geq 0,i+j=n} \lambda _{ij}\phi^i (x)
\psi ^j (x),
\ee
one obtains more general forms for the $n$--point functions. In this case,
generally the correlation function
$\langle{\cal O}(x_1){\cal O}(x_2)\cdots{\cal O}(x_n)\rangle$
has the ordinary scaling behaviour and
$\langle{\cal O}'(x_1)\cdots{\cal O}'(x_k){\cal O}(x_{k+1})\cdots{\cal O}(x_n)
\rangle$
has logarithmic behaviour up to $k$ logarithms.
Also, one can notice that the following relation between the correlation
functions are satisfied,
\be \label{DERIV}
\langle{\cal O}'(x_1)\cdots{\cal O}'(x_k){\cal O}(x_{k+1})\cdots{\cal O}(x_n)
\rangle=\frac{\partial ^k}{\partial \Delta ^k}
\langle{\cal O}(x_1){\cal O}(x_2)\cdots{\cal O}(x_n)
\rangle .
\ee
To show this, one must first {\it define} $\lambda_{ij}$'s ($i+j=n$) in terms
of the derivatives of $\lambda_n$, so that
\be
\sum_{i,j\geq 0,i+j=n}\lambda_{ij}\phi^i\psi^j={{\partial^n}\over
{\partial\DD^n}}\left({{\lambda_n}\over{n!}}\phi^n\right) ,
\ee
and then use this definition in the r. h. s. of (\ref{DERIV}).

The procedure presented above is an example of a general procedure to build
LCFT's from theories on AdS. One differentiates the action of the theory on
AdS, with respect to the weight (or one of the weights). The new theory on
AdS corresponds to an LCFT on $\partial$AdS, the correlators of which are
derivatives of the correlators of the former CFT with respect to the weight.



\begin{thebibliography}{99}
\bibitem{Ma} J. Maldacena; {\it The large N limit of superconformal field
             theories and supergravity}, hep-th/9711200.
\bibitem{GKP} S. Gubser, I. R. Klebanov, and A. M. Polyakov; {\it Gauge
              theory correlators from noncritical string theory},
              hep-th/9802109.
\bibitem{FF} S. Ferrera, and C. Fronsdal; {\it Gauge fields as composite
             boundary excitations}, hep-th/9802126.
\bibitem{W} E. Witten; {\it Anti de Sitter space and holography},
            hep-th/9802150.
\bibitem{MV} W. Muck, K. S. Viswanthan; {\it Conformal field theory
             correlators from classical field theory on AdS$_{d+1}$},
             hep-th/9804035.
\bibitem{HS} M. Henningson, K. Sfetsos; {\it Spinors and the AdS/CFT
             correspondence}, hep-th/9803251.
\bibitem{GKPF} A.M. Ghezelbash, K. Kaviani, S. Parvizi, A.H. Fatollahi;
              {\it Interacting spinors-scalars and AdS/CFT correspondence},
               hep-th/9805162, to appear in Phys. Lett. {\bf B}.
\bibitem{Gu} V. Gurarie; Nucl. Phys. {\bf B410}[FS] (1993) 535.
\bibitem{Sa} H. Saleur; Yale Preprint YCTP-P38-91, 1991.
\bibitem{RS} L. Rozansky and H. Saleur; Nucl. Phys. {\bf B376} (1992) 461.
\bibitem{KM} I.I. Kogan, N.E. Mavromatos; Phys. Lett.{\bf B375} (1996) 111.
\bibitem{CKLT} J.S. Caux, I.I. Kogan, A. Lewis and A.M. Tsvelik; Nucl. Phys.
           {\bf B489} (1997) 469.
\bibitem{KL} I.I. Kogan, A. Lewis, hep-th/9705240.
\bibitem{Ca} J. Cardy; J. Phys. A, 25 (1992) L201.
\bibitem{GFN} V. Gurarie, M.A.I. Flohr and C. Nayak, cond-mat/9701212.
\bibitem{WWH}  X.G. Wen, Y.S. Wu and Y. Hatsugai; Nucl. Phys. {\bf B422}[FS]
            (1994) 476.
\bibitem{rr1} M.R. Rahimi Tabar and S. Rouhani; Annals of Phys. {\bf 246}
            (1996)446.
\bibitem{rr2} M.R. Rahimi Tabar and S. Rouhani; Nouvo Cimento {\bf B112}
              (1997) 1079.
\bibitem{rr3} M.R. Rahimi Tabar and  S. Rouhani; Europhys. Lett. {\bf 37}
              (1997) 447.
\bibitem{flo} M.A. I. Flohr; Nucl. Phys. {\bf B482} (1996) 567.
\bibitem{rahim1} M.R. Rahimi Tabar and  S. Rouhani; Phys. Lett. {\bf A224 }
                (1997) 331.
\bibitem{GK} M.R. Gaberdiel, H.G. Kausch; Nucl. Phys. {\bf B447} (1996)
              293.
\bibitem{fl} M.A. I. Flohr; Int. J. Mod. Phys. {\bf A11} (1996) 4147.
\bibitem{BK} A. Bilal and I.I. Kogan; Nucl. Phys. {\bf B449} (1995) 569.
\bibitem{CKT} J.S. Caux, I.I. Kogan and A.M. Tsvelik; Nucl. Phys. {\bf B466}
           (1996) 444.
\bibitem{MS} Z. Maassarani, D. Serban;  Nucl. Phys. {\bf B489} (1997) 603.
\bibitem{MAG} M. Khorrami, A. Aghamohammadi, A.M. Ghezelbash;
{\it Logarithmic $N=1$ superconformal field theories}, hep-th/9803071, to appear
in
Phys. Lett. {\bf B}.
\bibitem{GhK} A.M. Ghezelbash, V. Karimipour; Phys. Lett. {\bf B402}
              (1997) 282.
\bibitem{RAK} M.R. Rahimi Tabar, A. Aghamohammadi, M. Khorrami; Nucl. Phys.
            {\bf B497} (1997) 555.
\bibitem{KAR} M. Khorrami, A. Aghamohammadi, M.R. Rahimi Tabar;
              Phys. Lett. {\bf B419} (1998) 179.
\bibitem{KA} M. Khorrami, A. Aghamohammadi; {\it Derivation of theories:
             structures of the derived system in terms of those of the
             original system in classical mechanics}, To appear in
             Il Nuovo Cimento {\bf B} (1998).
\bibitem{KAA} M. Khorrami, A. Aghamohammadi, M. Alimohammadi; {\it Derivation
              quantum theories: symmetries and the exact solution of the
              derived system}, to appear in Int.
              J. Mod. Phys. A (1998).
\end{thebibliography}
\end{document}